\documentclass[twocolumn]{revtex4}[12pt]
\draft
\usepackage{graphicx}
\begin{document}
\title{Langmuir dark solitons in dense ultrarelativistic electron-positron gravito-plasma in pulsar magnetosphere }
\author{U. A. Mofiz$^{1}$\footnote{Corresponding author; $\:$ email address: mofiz@bracu.ac.bd telephone:880-2-8824051-4 ext.4078, fax: 880-2-8810383.}   and M. R. Amin$^2$}
\address{$^1$Department of Mathematics and Natural Sciences, BRAC University, 66 Mohakhali, Dhaka-1212, Bangladesh\\
$^2$Department of Electronics and Communication Engineering, East West University, Jahurul Islam City. Aftabnagar, Dhaka-1212, Bangladesh\\
%$^2$Institut f\"{u}r Theoretische Physik IV, Fak\"{u}ltat f\"{u}r Physik and Astronomie, Ruhr-Universit\"{a}t Bochum, D-44780 Bochum, Germany
}

\vspace{1in}
\begin{abstract}

{\bf{Abstract}}\\
Nonlinear propagation of electrostatic  modes in ultrarelativistic dense elelectron-positron gravito-plasma at the polar cap region of pulsar magnetosphere is considered. A nonlinear Schr\"{o}dinger  equation is obtained from the reductive perturbation method which predicts the existence of Langmuir dark solitons.  Relevance of the propagating   dark solitons to the pulsar radio emission is discussed. 

\end{abstract}
%\bf{Keywords:} pulsar magnetosphere, pair plasma, electrostatic mode.
%\pacs{94.05.Fg; 94.05.-a; 52.35.}
%\pacs{PACS number(s): 52.35.Fp; 52.25.Zb; 52.35Mw; 52.35Sb; 52.35Tc}
%\begin{keywords}
%Second harmonic generation, laser radiation, ponderomotive force, piezoelectric semiconductor, Bohm potential.
%\end{keywords}
\maketitle
%\begin{sloppypar}
%\section*{I. INTRODUCTION}

\section*{I.  Introduction}
Pulsars are celestial sources that believe to be rotating neutron stars producing light-house like beams of radio emissions from the magnetic poles. As shown by Goldreich and Julian (1969) the rotating magnetic dipole produces a quadrupole electric field whose component parallel to the open magnetic field lines at the poles extracts particles very effectively from neutron star surface and accelerates them to highly relativistic energies. Thus, the magnetosphere is filled with plasma which shields the electric field.  Complete shielding is established when the net charge reaches $n_{GJ}$ - the Goldreich -Julian charge density. The Lorentz factors of the accelerated particles reach about $10^6$  and they emit hard curvature radiations that propagate at a sufficient angle to the magnetic field, so that significant pair production of electron-positron can occur (Erber 1966).  It is commonly accepted that the newly created particles produce more pairs by emitting energetic synchrotron or curvature radiation. As a result an avalanche of secondary particles populates the magnetosphere with densities $10^4n_{GJ}$ (Ruderman and Sutherland, 1975) .  Here, we extend our earlier research on pulsar microstructure, soliton formation, wakefield accelerations, gravitational waves ,  and growing modes (Mofiz, et al. 1985-2011) to account the pair ultrarelativistic  pressure and the gravity

The paper is organized as follows. Section II describes the fluid model of the dense ultrarelativistic electron-positron plasma under gravity. Considering a Lorentz invariant frame moving with group velocity of the wave, a Nonlinear  Schr\"{o}dinger Equation (NLSE) is derived using the reductive perturbation method (Gardner and Morikawa, 1960). A linear dispersion relation is obtained showing the existence of Langmuir waves under gravity and with ultrarelativistic temperature for wave propagation. The solution of NLSE shows the generation of  Langmuir dark solitons. Results are discussed in Sec. III. Finally, Sec. IV concludes the paper.

\section*{II.  The Mathematical Model}
We consider two-fluid magnetohydrodynamic (MHD) equations to describe the electron-positron plasma in the pulsar magnetosphere of the neutron star. The equations are the usual continuity and momentum balance equations for the plasma species, electrons and positrons, supplemented by the Poisson's equation for electrostatic wave propagation . Thus, the required set of equations are as follows (Mofiz and Ahmedov, 2000):
\begin{eqnarray}
\frac{\partial}{\partial t}\left(\gamma_sn_s\right)+\nabla\cdot\left(\sqrt{1-\frac{r_g}{r}} \gamma_sn_s\bf{v}_s\right)=0,\label{beq1}
\end{eqnarray}
\begin{eqnarray}
\frac{\partial}{\partial t}\left(\gamma_s\bf{v}_s\right) &+& \sqrt{1-\frac{r_g}{r}}\:\bf{v}_s\cdot\nabla\left(\gamma_s\bf{v}_s\right) \nonumber\\
&=& -\frac{q_s}{m}\nabla\phi-\frac{1}{m\gamma_sn_s}\nabla p_s,\:\:\:\: \label{beq2}
\end{eqnarray}
\begin{eqnarray}
\nabla\cdot\left(\frac{1}{\sqrt{1-\frac{r_g}{r}}}\nabla \phi\right)=-4\pi\sum_s \gamma_sq_s n_s, \label{beq3}
\end{eqnarray}
where, $\gamma_s=1/\sqrt{1-v_s^2/c^2}$, $n_s$, $\bf{v}_s$, and $q_s$ are respectively the particle number density, particle velocity, and particle charge of plasma species $s$; $q_s=-e$ for $s=e$ (electron) and $q_s=+e$ for $s=e^+$ (positron); $\phi$ is the electrostatic potential, $r_g$ is the Schwarzschild radius of the neutron star; $m$ is the electron/positron mass; $e$ is the absolute value of the electronic charge. In Eq. (2), the pressure $p_s$ is given by the expression for the ultrarelativistic pressure (Chandrasekhar, 1938): $p_s=n_0k_BT\left(n_s/n_0\right)^{4/3}$, where $T_e=T_p=T$ has been assumed; $k_B$ is the Boltzmann constant, $n_0$ is the equilibrium particle number density.

In the polar cap region of the pulsar, we consider $\theta=0$, $\nabla=\widehat{z}\partial/\partial z$, ${\bf{v}}_s=v_{sz}\widehat{z}$ and adopt the following normalization of different quantities: $z\rightarrow z\omega_{pe}/c$, $t\rightarrow t\omega_{pe}$, $n_s\rightarrow n_s/n_0$, $u_s\rightarrow v_{sz}/c$, $\phi\rightarrow e\phi/mc^2$, $\sigma_T\rightarrow k_BT/mc^2$, $r_g\rightarrow r_g\omega_{pe}/c$, and $v_g\rightarrow v_g/c$, where $\omega_{pe}=\left(4\pi n_0 e^2/m\right)^{1/2}$ is the electron plasma frequency, $c$ is the speed of light. To study the nonlinear dynamics, we consider the following stretched coordinates in the moving frame (Melikidze et al. 2000):
\begin{eqnarray}
\xi=\epsilon\gamma_0\left(z-v_gt\right),\label{beq4}
\end{eqnarray}
and
\begin{eqnarray}
\tau=\epsilon^2\gamma_0\left(t-v_gz\right),\label{beq5}
\end{eqnarray}
where, $v_g$ is the group velocity, $\gamma_0=\gamma_e=\gamma_p$ is the average relativistic Lorentz factor and is given by the following expression: $\gamma_0=\left(1-v_g^2\right)^{-1/2}$, $\epsilon$ is a small quantity,  the perturbation parameter with $\epsilon<1$. With the transformations, given by Eqs. (4) and (5), the derivatives $\partial/\partial t$ and $\partial/\partial z$ transform to $\partial/\partial t \rightarrow \partial/\partial t-\epsilon \gamma_0 v_g \partial/\partial \xi+\epsilon^2 \gamma_0\partial/\partial \tau$ and $\partial/\partial z \rightarrow \partial/\partial z+\epsilon \gamma_0 \partial/\partial \xi-\epsilon^2 \gamma_0 v_g\partial/\partial \tau$ respectively.
With the above considerations, the set of equations, Eqs. (1)-(3) take the following forms:
\begin{eqnarray}
\frac{\partial n_s}{\partial t} &+& \frac{\partial}{\partial z}\left(g(z)n_su_s\right)\nonumber\\
&+& \epsilon\left[-\gamma_0v_g\frac{\partial n_s}{\partial\xi}+\gamma_0\frac{\partial}{\partial\xi}\left(g(z)n_su_s\right)\right]\nonumber\\
&+& \epsilon^2\left[\gamma_0\frac{\partial n_s}{\partial\tau}-\gamma_0v_g\frac{\partial}{\partial\tau}\left(g(z)n_su_s\right)\right] =0, \label{beq6}
\end{eqnarray}

\begin{eqnarray}
\frac{\partial u_s}{\partial t} &+& g(z)u_s\frac{\partial u_s}{\partial z}+\frac{q_s}{e\gamma_0}\frac{\partial\phi}{\partial z}+\frac{4\sigma_T}{3\gamma_0^2}n_s^{-2/3}\frac{\partial n_s}{\partial z}\nonumber\\
&+& \epsilon\left[-\gamma_0v_g\frac{\partial u_s}{\partial\xi}+g(z)u_s\gamma_0\frac{\partial u_s}{\partial\xi} \right.\nonumber\\
&+& \left. \frac{q_s}{e}\frac{\partial\phi}{\partial\xi}+\frac{4\sigma_T}{3\gamma_0}n_s^{-2/3}\frac{\partial n_s}{\partial\xi}\right]\nonumber\\
&+& \epsilon^2\left[\gamma_0\frac{\partial u_s}{\partial\tau}-g(z)\gamma_0v_gu_s\frac{\partial u_s}{\partial\tau} \right.\nonumber\\
&-& \left. \frac{q_sv_g}{e}\frac{\partial\phi}{\partial\tau}-\frac{4\sigma_Tv_g}{3\gamma_0}n_s^{-2/3}\frac{\partial n_s}{\partial\tau}\right] =0, \label{beq7}
\end{eqnarray}

\begin{eqnarray}
\frac{\partial}{\partial z}\left(\frac{1}{g(z)}\frac{\partial\phi}{\partial z}\right) &-& \gamma_0\left(n_e-n_p\right) \nonumber\\
&+& \epsilon\left[\gamma_0\frac{\partial}{\partial\xi}\left(\frac{1}{g(z)}\frac{\partial\phi}{\partial z}\right) \right.\nonumber\\
&+& \left.\frac{\partial}{\partial z}\left(\frac{\gamma_0}{g(z)}\frac{\partial\phi}{\partial\xi}\right)\right]\nonumber\\
&+& \epsilon^2\left[-\gamma_0v_g\frac{\partial}{\partial\tau}\left(\frac{1}{g(z)}\frac{\partial\phi}{\partial z}\right) \right.\nonumber\\
&-& \left. \gamma_0v_g\frac{\partial}{\partial z}\left(\frac{1}{g(z)}\frac{\partial\phi}{\partial\tau}\right) \right.\nonumber\\
&+& \left. \gamma_0\frac{\partial}{\partial\xi}\left(\frac{\gamma_0}{g(z)}\frac{\partial\phi}{\partial\xi}\right)\right]\nonumber\\
&+& \epsilon^3\left[-2\gamma_0^2\frac{\partial}{\partial\xi}\left(\frac{v_g}{g(z)}\frac{\partial\phi}{\partial\tau}\right)\right.\nonumber\\
&-& \left. \gamma_0^2v_g\frac{\partial}{\partial\tau}\left(\frac{1}{g(z)}\frac{\partial\phi}{\partial\xi}\right)\right]\nonumber\\
&+& \epsilon^4\left[\gamma_0^2v_g^2\frac{\partial}{\partial\tau}\left(\frac{1}{g(z)}\frac{\partial\phi}{\partial\tau}\right)\right]
=0, \label{beq8}
\end{eqnarray}
where, the factor $g(z)=\left(1-r_g/z\right)^{1/2}$ accounts for the gravitational effect. Now we expand the quantities $n_s$, $u_s$, $\phi$ as
\begin{eqnarray}
n_s &=& 1+\epsilon^2n_{s0} \nonumber\\
&+& \sum_{l=1}^{\infty} \epsilon^l\left(n_{sl}e^{il(kz-\omega t)}+n_{sl}^*e^{-il(kz-\omega t)}\right),
\label{beq9}
\end{eqnarray}
\begin{eqnarray}
u_s=\epsilon^2u_{s0}+\sum_{l=1}^{\infty} \epsilon^l\left(u_{sl}e^{il(kz-\omega t)}+u_{sl}^*e^{-il(kz-\omega t)}\right), \label{beq10}
\end{eqnarray}
\begin{eqnarray}
\phi=\epsilon^2\phi_0+\sum_{l=1}^{\infty} \epsilon^l\left(\phi_le^{il(kz-\omega t)}+\phi_l^*e^{-il(kz-\omega t)}\right). \label{beq11}
\end{eqnarray}
Here,\\
 $(n_{s0}$, $u_{s0}$, $\phi_0$ , $n_{sl}$, $u_{sl}$, $\phi_l)$ $\equiv$ $A^{(1)}+\epsilon A^{(2)}+\epsilon^2A^{(3)}+.......$
 are functions of stretched coordinates $(\xi,\tau)$.

\noindent { \bf{II. A. Linear Dispersion Relation for the Langmuir Wave}
\vspace{0.1in}}\\
   Now considering $ \left|1/g(z)\cdot dg(z)/dz\right| <<k$
 for the first harmonic ($l=1$)in the first order ($\epsilon=1$  ), we have the following equations for the first-order quantities:

\begin{eqnarray}
-i\omega n_{s1}^{(1)}+ikg(z)u_{s1}^{(1)}=0, \label{beq12}
\end{eqnarray}
\begin{eqnarray}
-i\omega u_{s1}^{(1)}+\frac{4ik\sigma_T}{3\gamma_0^2}n_{s1}^{(1)}+\frac{ikq_s}{e\gamma_0}\phi_1^{(1)}=0, \label{beq13}
\end{eqnarray}
\begin{eqnarray}
-\frac{k^2}{g(z)}\phi_1^{(1)}-\gamma_0\left(n_{e1}^{(1)}-n_{p1}^{(1)}\right)=0. \label{beq14}
\end{eqnarray}
Eliminating $u_{s1}^{(1)}$ from Eqs. (12) and (13), we obtain the following equation relating $n_{s1}^{(1)}$ and $\phi_1^{(1)}$:
\begin{eqnarray}
n_{s1}^{(1)}=-\frac{k^2g(z)}{-\omega^2+4\sigma_Tg(z)k^2/3\gamma_0^2}\frac{q_s}{e\gamma_0}\phi_1^{(1)}, \label{beq15}
\end{eqnarray}
from which we obtain
\begin{eqnarray}
n_{e1}^{(1)}-n_{p1}^{(1)}=\frac{2k^2g(z)}{-\omega^2+4\sigma_Tg(z)k^2/3\gamma_0^2}\frac{1}{\gamma_0}\phi_1^{(1)}. \label{beq16}
\end{eqnarray}
Using Eq. (16) into Eq. (14), we obtain the following linear dispersion relation:
\begin{eqnarray}
\omega^2=2g^2(z)+\frac{4\sigma_Tg(z)}{3\gamma_0^2}k^2, \label{beq17}
\end{eqnarray}
which in the dimensional form is
\begin{eqnarray}
\omega^2=\omega_{pe}^2g^2(z)+\frac{2}{3}k^2v_{th}^2g(z), \label{beq18}
\end{eqnarray}
with $v_{th}^2=\frac{2k_BT}{m\gamma_0^2}$.
The group velocity $v_g=\partial\omega/\partial k$ is obtained from the linear dispersion relation Eq. (17) as:
\begin{eqnarray}
v_g=\frac{4\sigma_Tg(z)}{3\gamma_0^2} \frac{k}{\omega}. \label{beq19}
\end{eqnarray}
The same expression for $v_g$ is also obtained from the compatibility condition and shown in the Appendix A.\\ \\
  The group dispersion is found to be
\begin{eqnarray}
v_g^{'}=\frac{dv_g}{dk}=\frac{\frac{4}{3}\frac{\sigma_Tg(z)}{\gamma_0^2}-v_g^2}{\omega}. \label{beq20}
\end{eqnarray}
Eq.(17) represents the dispersion relation for Langmuir waves in ultrarelativistic $e,e^+$ plasma under gravity. Pair production in the polar cap region of pulsar magnetosphere occurs through curvature radiation which happens for $\mathcal{E}_\parallel>>m_ec^2$, where $\mathcal{E}_\parallel$ is the energy of electron along the magnetic field. It is estimated that for cascade generation of pair plasma $\gamma_0\sim 10^6-10^7$,$\mathcal{E}_\parallel\sim 10^{12}-10^{13}eV$ (Beskin et al. 1993). Here, $\gamma_0=\mathcal{E}_\parallel/m_ec^2$,then considering $\mathcal{E}_\parallel=k_BT$, we find $\gamma_0=k_BT/m_ec^2\equiv\sigma_T$. Using Eqs.(17),(19) and Eq.(20), we perform an analysis of the dispersion relation, group velocity and group dispersion of Langmuir waves at the ultrarelativistic temperature of the $e,e^+$ plasma under gravity. The analysis is shown graphically
 in Fig.1-4, respectively.\\
\begin{figure}[ht!]% \vspace{6.50cm}
\centering
\includegraphics[height=6.5cm, width=8.5cm]{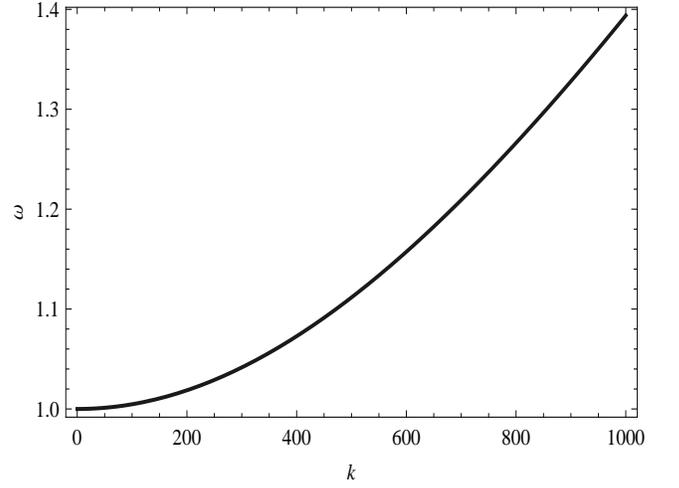}
%\special{psfile=fig1.eps}
\caption{The variation of the normalized frequency $\omega$ of the linear Langmuir wave with respect to the normalized pump wavenumber $k$ for different values of plasma parameters : $r_g=1$, $z=2$, $\gamma_0=\sigma_T=10^6$ with the corresponding temperature $T=5\times 10^{15}\: K$ .}
\end{figure}

\begin{figure}[ht!]% \vspace{6.50cm}
\centering
\includegraphics[height=6.5cm, width=8.5cm]{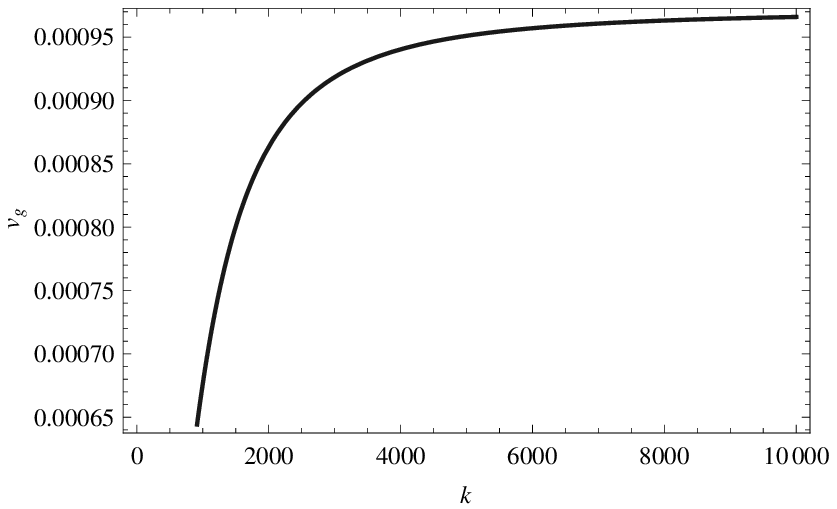}
%\includegraphics[height=6.5cm, width=8.5cm]{fig33.eps}
%\special{psfile=fig1.eps}
\caption{The variation of the normalized group velocity $v_g$ of the linear Langmuir wave with respect to the normalized pump wavenumber $k$ for different values of plasma parameters : $r_g=1$, $z=2$, $\gamma_0=\sigma_T=10^6$ with the corresponding temperature $T=5\times 10^{15}\: K$ .}
\end{figure}

\begin{figure}[ht!]% \vspace{6.50cm}
\centering
\includegraphics[height=6.5cm, width=8.5cm]{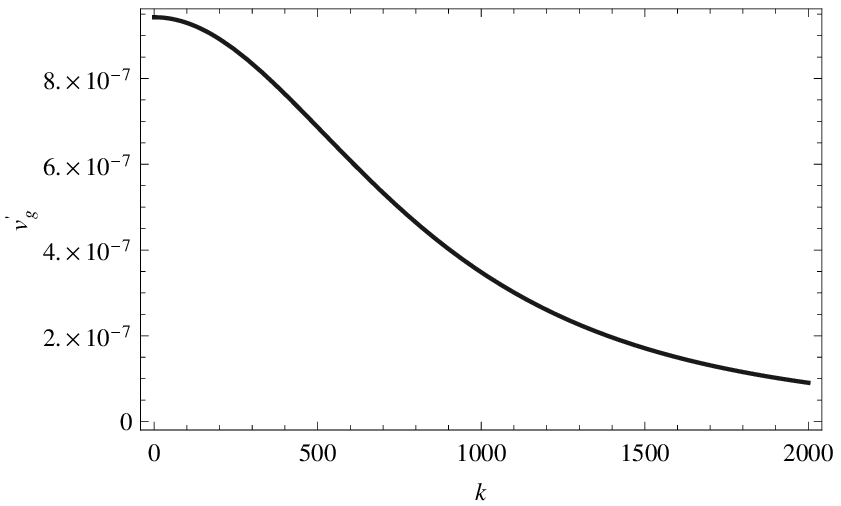}
%\includegraphics[height=6.5cm, width=8.5cm]{fig33.eps}
%\special{psfile=fig1.eps}
\caption{The variation of the normalized group dispersion $v_g'$ of the linear Langmuir wave with respect to the normalized pump wavenumber $k$ for different values of plasma parameters : $r_g=1$, $z=2$, $\gamma_0=\sigma_T=10^6$ with the corresponding temperature $T=5\times 10^{15}\: K$ .}
\end{figure}

\begin{figure}[ht!]% \vspace{6.50cm}
\centering
\includegraphics[height=6.5cm, width=8.5cm]{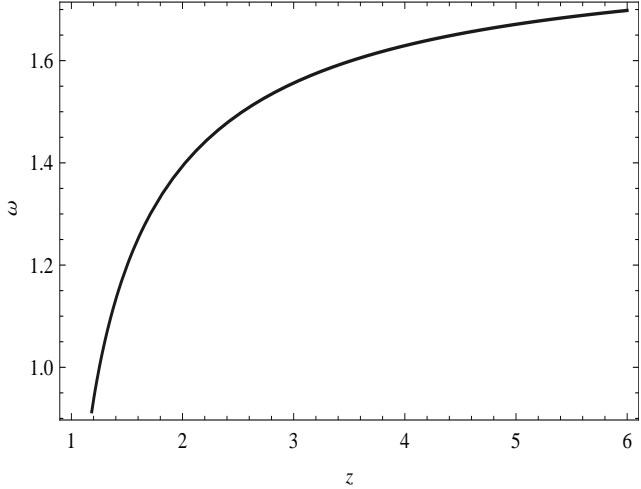}
%\includegraphics[height=6.5cm, width=8.5cm]{fig32.eps}
%\special{psfile=fig32.eps}
\caption{The variation of the normalized frequency $\omega$ of the linear Langmuir wave with respect to the normalized distance $z$ for  parameters : $r_g=1$, $k=1000$,$\gamma_0=\sigma_T=10^6$ with the corresponding temperature $T=5\times 10^{15}\: K$ . }
\end{figure}

\noindent {\bf{II.B Nonlinear Evolution Equation for the Langmuir Wave\vspace{0.1in}}}\\
Finding the zeroth harmonic   and second harmonic   of the second order   quantities in terms of the first harmonic   of the first order quantities   and using these into the first harmonic   of the third order quantities   , we easily obtain the following NLSE for the evolution of the potential
\begin{eqnarray}
i\frac{\partial a}{\partial\tau}+P\frac{\partial^2 a}{\partial\xi^2}
+Q\left|a\right|^2a=0, \label{beq21}
\end{eqnarray}
where $a\equiv \phi_1^{(1)}$, and the coefficients $P$ and $Q$ are given by the following expressions:
\begin{eqnarray}
P &=& \frac{\frac{4\sigma_Tg(z)}{3\gamma_0^2}-v_g^2}{\frac{2}{\gamma_0}\left[\omega-\frac{v_g}{kg^3(z)}\left\{2\omega^2+v_gk\omega
+2g^2(z)\right\}\right]}, \label{beq22}
\end{eqnarray}

\begin{eqnarray}
 Q &=& \frac{\frac{g^2(z)}{\gamma_0}\left[\frac{\omega}{k}f_{11}+h_{11}\right]}{\frac{2}{\gamma_0}\left[\omega-\frac{v_g}{kg^3(z)}\left\{2\omega^2+v_gk\omega
+2g^2(z)\right\}\right]}, \label{beq23}
\end{eqnarray}
where
\begin{eqnarray}
f_{11} &=& -\frac{\omega k}{g^2(z)\gamma_0}\left[b_1\left(1+\frac{kv_g}{\omega}\right)+2b_2\right]\nonumber\\
&+& \frac{3\omega k^5}{4g^4(z)\gamma_0^3}, \label{beq24}
\end{eqnarray}
\begin{eqnarray}
h_{11} &=& -\frac{\omega^2}{g^2(z)\gamma_0}\left[\frac{kv_g}{\omega}\frac{b_1}{g(z)}+b_2-\frac{3k^4}{4g^3(z)\gamma_0^2}
\right] \nonumber\\
&+& \frac{8\sigma_T k^2}{9g^2(z)\gamma_0^3}\left(b_1+b_2\right), \label{beq25}
\end{eqnarray}
with
\begin{eqnarray}
b_1 &=& -\frac{1}{\frac{4\sigma_Tg(z)}{3\gamma_0^2}-v_g^2}\left[g^2(z)\left(\frac{\omega k}{2g^2(z)\gamma_0}\right)^2\right.\nonumber\\
&-& \left. \frac{2\sigma_T k^4}{9g(z)\gamma_0}+\frac{v_g\omega k^3}{2g^2(z)\gamma_0^2}\right], \label{beq26}
\end{eqnarray}

\begin{eqnarray}
b_2 &=& \frac{3\omega^2 k^4}{16g^4(z)\gamma_0^2}-\frac{\sigma_T k^4}{9g^2(z)\gamma_0^3}. \label{beq27}
\end{eqnarray}
Here, the coefficient $P$ can be written as
\begin{eqnarray}
P = \frac{1}{2\alpha} v_g^{'}, \label{beq28}
\end{eqnarray}
where,
%\begin{eqnarray}
%\alpha = \frac{1}{\gamma_0}[1-\frac{v_g}{\omega k g^3(z)}\{3\omega^2+\omega(kv_g-\omega v_g^2)\}], \label{beq29}
%\end{eqnarray}

\begin{eqnarray}
\alpha = \frac{1}{\gamma_0}[1-\frac{4\sigma_T}{\gamma_0^2g^2(z)} ], \label{beq29}
\end{eqnarray}

represents the effects of ultrarelativistic temperature and gravity , neglecting of which we recover the usual results.
The coefficients $P$ and $Q$ appearing in the NLSE, given by Eqs. (22), (23)are known as the dispersion and nonlinear coefficients, respectively. The signs of $P$ and $Q$ determine whether the slowly varying wave pulse is stable or not (Lighthill condition; Lighthill, 1967). If the signs of $P$ and $Q$ are such that $PQ<0$, the wave pulse is modulationally stable and the corresponding solution of the NLSE is called the dark soliton. On the other hand, if $PQ>0$, then the pulse may be modulationally unstable and the solution of the NLSE in this case is called the bright soliton. Graphically, we study the  nature of $P$ and $Q$ for continuous values of the wave number $k$ with particular values of plasma parameters, which are shown graphically in Fig.5-6,  respectively.

\begin{figure}[ht!]% \vspace{6.50cm}
\centering
\includegraphics[height=6.5cm, width=8.5cm]{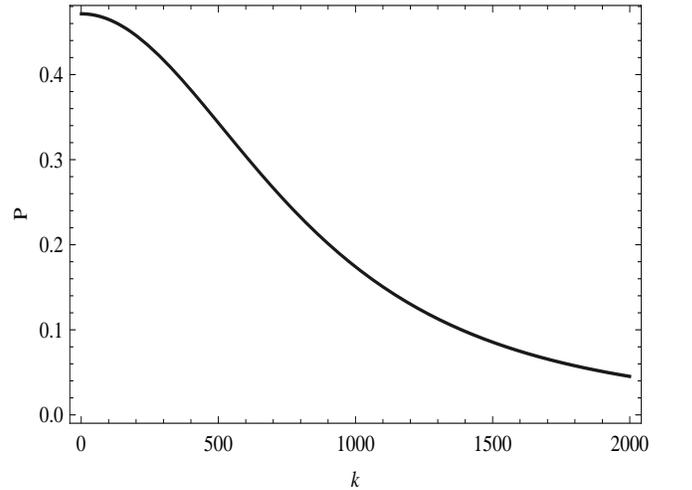}
%\includegraphics[height=6.5cm, width=8.5cm]{fig34.eps}
%\special{psfile=fig34.eps}
\caption{The variation of the dispersion coefficient $P$ of the linear Langmuir wave with respect to the normalized pump wavenumber $k$ for different values of plasma parameters : $r_g=1$, $z=2$, $\gamma_0=\sigma_T=10^6$ with the corresponding temperature $T=5\times 10^{15}\: K$ .}
\end{figure}

\begin{figure}[ht!]% \vspace{6.50cm}
\centering
\includegraphics[height=6.5cm, width=8.5cm]{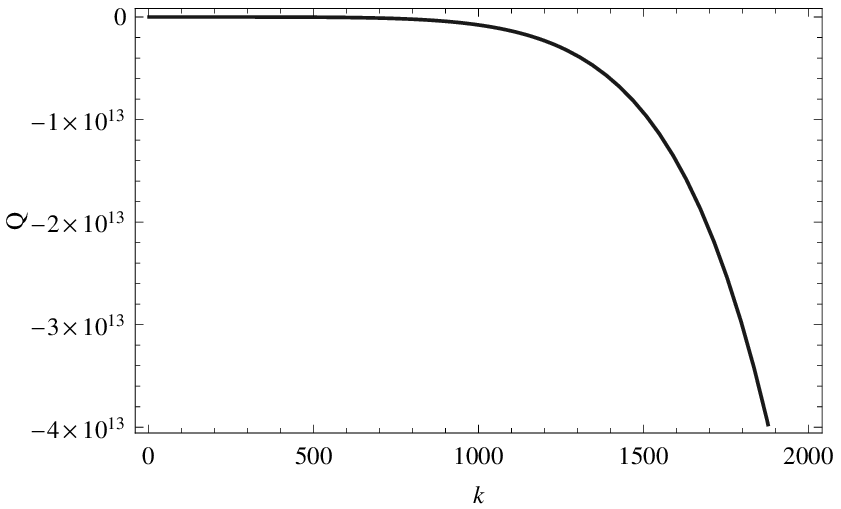}
%\includegraphics[height=6.5cm, width=8.5cm]{fig35.eps}
%\special{psfile=fig35.eps}
\caption{The variation of the nonlinear coefficient $Q$ of the linear Langmuir wave with respect to the normalized pump wavenumber $k$ for different values of plasma parameters : $r_g=1$, $z=2$, $\gamma_0=\sigma_T=10^6$ with the corresponding temperature $T=5\times 10^{15}\: K$ .}
\end{figure}

From the graphical analysis, we find that $PQ<0$. Then applying the standard technique, and by taking $a(\xi,\tau)=a(\xi)\:\exp{[i(K\xi-\Omega\tau)]}$, the following solution of the NLSE, Eq. (21) (Mofiz,2007)is easily obtained:
\begin{eqnarray}
a(\xi, \tau) &=& a_0\; \mbox{tanh}\left[\left|\frac{Q}{4P}\right|^{1/2}a_0\xi\right]\exp{[i(K\xi-\Omega\tau)]}, \label{beq30}
\end{eqnarray}
Here,$K\xi-\Omega\tau$ is the modulation phase with $K(<<k)$ and $\Omega(<<\omega)$, respectively.
Eq.(30) represents a dark soliton with amplitude\\
$a_0=\left|\frac{\Omega+PK^2}{Q}\right|^{1/2}$ and width $\delta=\left|\frac{4P}{Qa_0^2}\right|^{1/2}$, respectively. The dark soliton (Eq.(30)) in the ultrarelativistic $e,e+$ plasma is  shown graphically in Fig.7. \\
%\begin{figure}[ht!]% \vspace{6.50cm}
%\centering
%\includegraphics[height=6.5cm, width=8.5cm]{FIG3.7.eps}
%\special{psfile=fig37.eps}
%\caption{Dark soliton in ultrarelativistic $e,e^+$ plasma at the polar cap region of pulsar magnetosphere. The parameters are : $r_g=1$, $z=2$, $k=1000$,$\Omega=0$,$K=1$,, $\gamma_0=\sigma_T=10^6$ with the corresponding temperature $T=5\times 10^{15}\: K$ .}
%\end{figure}

\begin{figure}[ht!]% \vspace{6.50cm}
\centering
\includegraphics[height=6.5cm, width=8.5cm]{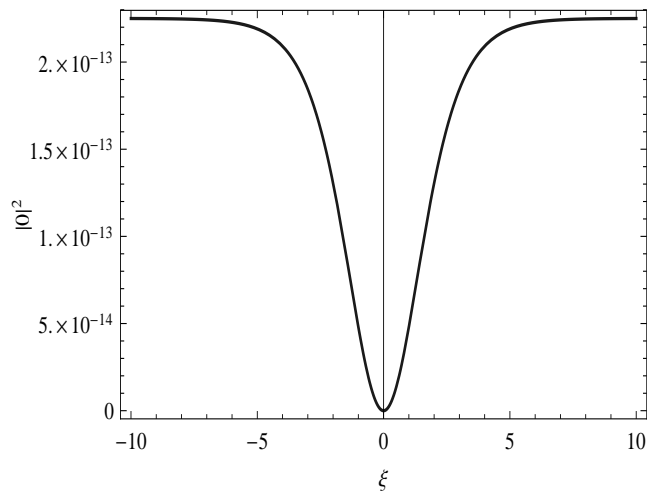}
%\special{psfile=fig37.eps}
\caption{Dark soliton in ultrarelativistic $e,e^+$ plasma at the polar cap region of pulsar magnetosphere. The parameters are : $r_g=1$, $z=2$, $k=1000$,$\Omega=0$,$K=1$,, $\gamma_0=\sigma_T=10^6$ with the corresponding temperature $T=5\times 10^{15}\: K$ .}
\end{figure}

\section*{III.  Results and Discussion}

In this section, we analyze the linear dispersion as well as the nonlinear Langmuir dark soliton in the pulsar magnetosphere. 
Eq.(17) shows that the Langmuir frequency depends on ultrarelativistic temperature and it  is redshifted due to gravity near the Schwarzchild radius. Similarly, the group velocity and group dispersion, shown by Eq.(19) and Eq. (20), are also depend on temperature and gravity.

For numerical appreciation of the dark soliton ,we consider the two cases of ultrarelativistic temperatures: $6\times 10^{10}K-1.8\times 10^{11}K$ (Crab pulsar) with the corresponding energies $5-15 MeV$ (Nanobashvilli,2004) and $5\times 10^{15}K-5\times 10^{16}K$ (x-ray pulsar) with the corresponding energies $10^{12}-10^{13} eV$ (Beskin et al., 1993).
 
The solution of the NLSE (Eq.(21)) is a stable dark soliton (Eq.(30))whose amplitude and width depend on temperature. The amplitude is increased and the width is decreased with the increase of ultrarelativistic temperature. Thus, stable spiky Langmuir solitons are possible in the ultrarelativistic electron-positron plasma.\\

%$\mathcal{E}_\parallel$

\section*{IV.  Conclusion}

To summarize, we have investigated the nonlinear propagation of electrostatic modes in a dense ultrarelativistic electron-positron gravito-plasma at the polar cap region of pulsar magnetosphere. A multiscale perturbation analysis of the fluid equations shows that  stable dark Langmuir solitons are produced  due to the balnce of dispersion and nonlinearity in the wave propagation. As the amplitude of the soliton increaes and width of the soliton decreases with the increase of ultrarelativistic temperature, so spiky stable dark Langmuir solitons may propagate along the open field lines of the pulsar magnetosphere, which may have some relation with pulsar radio emission and its  microstructure.

\section*{Acknowledgement}
This work has been supported by the Ministry of Education of the Government of Bangladesh under Grants for Advanced Research in Science: MOE.ARS.PS.2011. No.-86.

\section*{Appendix A: The compatibility condition}

%\noindent
%{\bf Appendix A}\\

It can be shown that the 1st harmonic of the 2nd-order electron and positron densities can be found to be
$$
n_{s1}^{(2)}=\frac{q_s k^2}{2e g^2(z)\gamma_0}\phi_1^{(2)}+\frac{i\gamma_0}{2g^2(z)} $$
$$\times \left[\left(\omega v_g-\frac{4\sigma_T g(z) k}{3\gamma_0^2}\right)\frac{\partial n_{s1}^{(1)}}{\partial\xi}+g(z)\left(kv_g-\omega\right)\frac{\partial u_{s1}^{(1)}}{\partial\xi}\right] $$
$$-\frac{iq_s k}{2eg(z)}\frac{\partial \phi_1^{(1)}}{\partial\xi}.
$$

After finding $n_{e1}^{(2)}-n_{p1}^{(2)}$ and substituting it in the following 1st-harmonic of the 2nd-order part of the Poisson's equation:
$$
-\frac{k^2}{g(z)}\phi_1^{(2)}-\gamma_0\left(n_{e1}^{(2)}-n_{p1}^{(2)}\right)+\frac{2ik\gamma_0}{g(z)}\frac{\partial \phi_1^{(1)}}{\partial\xi}=0,
$$
we obtain the following compatibility condition:
$$
v_g=\frac{4\sigma_T g(z)}{3\gamma_0^2}\frac{k}{\omega},
$$
which is exactly the same as the expression of the group velocity, Eq. (19), obtained by differentiating $\omega$ with respect to $k$ from the linear dispersion relation, Eq. (17).

\end{document}